\documentclass[useAMS,usenatbib]{mn2e}
\usepackage{psfig,graphicx}
\newcommand{\ee}{\end{equation}}
\newcommand{\bea}{\begin{eqnarray}}
\newcommand{\eea}{\end{eqnarray}}

\newcommand{\bfig}{\begin{figure}}
\newcommand{\efig}{\end{figure}}
\newcommand{\bfigw}{\begin{figure*}}
\newcommand{\efigw}{\end{figure*}}
\newcommand{\bmp}{\begin{minipage}}
\newcommand{\emp}{\end{minipage}}
\newcommand{\etal}{et al.~}

\newcommand{\eg}{{e.g.}}
\newcommand{\ie}{{i.e. }}
\newcommand{\bpic}{\begin{picture}}
\newcommand{\epic}{\end{picture}}

\newcommand{\ROSAT}{{\em ROSAT }}

\newcommand{\PSPC}{{\em PSPC }}

\newcommand{\Chandra}{{\em Chandra }}

\newcommand{\XMMc}{{\em XMM-Newton}}

\newcommand{\IRAS}{{\em IRAS }}

\newcommand{\cm}{{\rm\thinspace cm}}

\newcommand{\erg}{{\rm\thinspace erg}}
\newcommand{\Jy}{{\rm\thinspace Jy}}
\newcommand{\mJy}{{\rm\thinspace mJy}}
\newcommand{\uJy}{{\rm\thinspace \mu Jy}}

\newcommand{\MHz}{{\rm\thinspace MHz}}
\newcommand{\GHz}{{\rm\thinspace GHz}}

\newcommand{\keV}{\hbox{{\rm\thinspace keV}}}

\newcommand{\km}{{\rm\thinspace km}}

\newcommand{\Mpc}{{\rm\thinspace Mpc}}

\newcommand{\s}{{\rm\thinspace s}}

\newcommand{\fluxerg}{\hbox{$\erg\cm^{-2}\s^{-1}\,$}}

\newcommand{\ergps}{\mbox{$\erg\s^{-1}$}}

\newcommand{\kmpspMpc}{\hbox{$\km\s^{-1}\Mpc^{-1}\,$}}

\newcommand{\um}{\hbox{$\mu {\rm m}$}}

\newcommand{\rband}{$R$--\,band~}

\newcommand{\uv}{{\it uv }}

\newcommand{\kl}{k\lambda}
\newcommand{\aips}{{\sc aips }}
\newcommand{\imagr}{{\sc imagr }}
\newcommand{\calib}{{\sc calib }}
\newcommand{\clean}{{\sc clean }}
\newcommand{\dbcon}{{\sc dbcon }}
\newcommand{\sad}{{\sc sad }}
\newcommand{\immod}{{\sc immod }}

\newcommand{\pbcorc}{{\sc pbcor}}

\newcommand{\aipsc}{{\sc aips}}
\newcommand{\imagrc}{{\sc imagr}}

\newcommand{\dbconc}{{\sc dbcon}}
\newcommand{\sadc}{{\sc sad}}

\newcommand{\mch}{M$\rm^{c}$Hardy }
\newcommand{\mchc}{M$\rm^{c}$Hardy} 
\title[Radio Observations of the 13hr {\it XMM/ROSAT} Deep X-ray Survey Area]{Radio Observations of the 13hr {\it XMM/ROSAT} Deep X-ray Survey Area}
\author[N. Seymour, I.M. \mch \& K.F. Gunn]{N. Seymour,$^{1,2,}$\thanks{E-mail: seymour@iap.fr} I.M. \mchc$^2$ \& K.F. Gunn$^2$\\
$^{1}$Institut d'Astrophysique de Paris, 98bis. Boulevard Arago, 75014 Paris, France \\ 
$^{2}$Department of Physics and Astronomy, University of Southampton, Highfield, Southampton, SO17 IBJ, UK}

\begin{document}

\date{DRAFT VERSION}

\pagerange{\pageref{firstpage}--\pageref{lastpage}} \pubyear{2002}

\maketitle

\label{firstpage}

\begin{abstract}
In order to determine the relationship between the faint X-ray and
faint radio source populations, and hence to help understand the X-ray
and radio emission mechanisms in those faint source populations, we
have made a deep $1.4\GHz$ Very Large Array radio survey of the $13^h$
$+38^{\circ}$ {\it XMM/ROSAT} X-ray Survey Area. From a combined data set of
10hours B configuration data and 14hours A configuration data, maps
with $3.35''$ resolution and a noise limit of $7.5\uJy$ were
constructed.  A complete sample of 449 sources was detected within a
$30'$ diameter region above a $4\sigma$ detection limit of $30\uJy$, at the map
centre, making this one of the deepest radio surveys at this
frequency. The differential source count shows a significant upturn at
sub-milliJansky flux densities, similar to that seen in other deep surveys at
$1.4\GHz$ (\eg\space the Phoenix survey), but larger than that seen in the HDF
which may have been selected to be under-dense. This upturn is well modelled 
by the emergence of a population of medium redshift starforming 
galaxies which dominate at faint flux densities. The brighter source counts are
well modelled by AGNs. 
\end{abstract}


\section{Introduction}

In recent years, deep radio surveys \citep[\eg][etc.]
{windhorst90,condon89,prandoni01,hdf,pdfs,bondi03,hopkins03,gruppioni99,ciliegi99}
have shown that there is an upturn in the differential radio source
counts, normalised to a Euclidean value, below $\sim 1\mJy$. This
upturn has generally been interpreted as the emergence of a new
population of starforming galaxies \citep[\eg][]{condon89,rr93,garrett01} 
which does not show up at higher flux densities, where number counts are 
dominated by AGN. Given the strong
correlation between $1.4\GHz$ radio flux and the far infrared flux
\citep{hsrr85,carilliyun00}, which is itself a starformation
rate indicator, radio luminosities can now be used as an independent measure 
of the starformation rate and, combined with other data, as a way of 
deconvolving the contribution of AGN and starbursts to the energy budget 
of the universe.

In order to investigate the nature of the objects that contribute to
the X-ray Background radiation (XRB) a deep X-ray survey \citep{imh98} 
with \ROSAT was performed, using the \PSPC detector, of a field at RA
13 34 37 Dec 37 54 44 (J2000). 96 sources were detected in a $30'$
diameter field above a flux limit of $2\times 10^{-15}\fluxerg$ in the
$0.5-2.0\keV$ band. The majority of the identifications were with
broad line AGN, but some galaxies with narrow optical
emission lines, known as narrow emission line galaxies (NELGs), 
were found. 
These galaxies had a range of X-ray luminosities from $\sim10^{40}$ to
$\sim10^{43} \ergps$, and a wide range of X-ray/optical ratios and
optical emission line ratios. The highest X-ray luminosity yet found
for a starburst galaxy is $\sim10^{42} \ergps$
\citep{moran99,halpern95} and so the higher 
luminosity NELGs are most likely to be obscured AGN. However the lower
luminosity NELGs, with low ($\sim0.01 - 0.001$) X-ray/optical ratios
could easily be explained by starburst emission
\citep[\eg][]{kfg01}. Our conclusion was that the X-ray emission from NELGs, 
both as a sample and within individual galaxies, is due to a mixture 
of AGN and starbursts.

As a starburst is associated with radio emission
\citep[\eg][]{normalgalaxies} then, if some of the NELG X-ray emission is
produced by starbursts, one would expect detectable, associated radio emission; 
steep radio spectrum emission, detected and resolved on galaxy scales, 
would be a very strong starburst indicator \citep[\eg][]{normalgalaxies}.
Preliminary $1.4\GHz$ observations made with the
Very Large Array (VLA) radio telescope of the US National Radio
Astronomy Observatory\footnote{The National Radio Astronomy
Observatory is a facility of the National Science Foundation operated
under cooperative agreement by Associated Universities, Inc.} were 
reported in \cite{imh98}. We noted that 4 of the 10 brightest NELGs were 
detected as radio sources whereas a much lower fraction of X-ray QSOs was 
detected indicating that complementary radio observations would help 
determine the emission processes of these NELGs. 

In order to improve on the significance of this early result we
carried out deeper observations with the VLA at $1.4\GHz$, and at $5\GHz$, of
the $30'$ diameter X-ray survey area. At $1.4\GHz$ the primary beam diameter
of the VLA is $30'$ and so the whole X-ray field can be covered in a
single pointing. Thus the X-ray and $1.4\GHz$ radio fields are very well
matched. At $5\GHz$, due to the smaller size of the primary beam, we have 
to make a number of pointings, over a grid of positions, in order to fully 
cover the X-ray field. We also 
performed higher resolution ($\sim 0.2''$ beam) observations with
MERLIN at $1.4\GHz$. The ultimate aim is to resolve and detect any extended 
radio emission on galaxy scales out to redshifts $z>0.5$ and to determine
the radio spectra of the various sources. 

X-ray spectra are also a strong diagnostic of the emission mechanism,
\eg\space soft thermal spectra would be associated with starburst emission.
With \XMMc, we therefore made a very deep (200ksec) observation
(Page et al. 2003; Loaring et al. 2004, in preparation), providing
high quality X-ray spectra in the $0.1-12 \keV$ band, and detecting
fainter X-ray sources than are seen in the \ROSAT survey. In order to
provide X-ray positions of sub arcsecond accuracy, and hence
unambiguous identifications, we also observed the 13hr field for $4
\times 30$ksec with the \Chandra ACIS-I instrument \citep{chandracat}.
We detected 214 X-ray sources.  A preliminary analysis of the
X-ray/radio relationship, based on these new data, is given by
\cite{gunn03}.

In this paper we present the new $1.4\GHz$ VLA observations. The $5\GHz$ 
observations are described in Seymour 2002\nocite{seymour02}, and the 
MERLIN observations and a detailed correlation of the X-ray and radio 
sources will be presented elsewhere. The 
observations and data reduction are described in Section 2. We note that 
our observational set-up and analysis techniques are
similar to those of \cite{hdf}, to whom we refer readers for a
more detailed description. The source
extraction is described in Section 3. This $1.4\GHz$ survey is
comparable in depth to the deepest surveys previously made at this
frequency \citep[\eg][and references therein]{hdf,hopkins03} and we
detect sources at $4\sigma$ significance down to 
$30\uJy$ at the map centre. We are therefore able to define
accurately the shape of the source counts below $1\mJy$. In Section 4
we present these source counts, compare them with previous results and model 
them. The radio/optical/X-ray cross-correlation will be 
presented in future papers.

\section{Observations and Data Reduction}
\label{sec:radobs}

In November 1995 the 13hr field was observed for
10hours at $1.4\GHz$ by the VLA in the B configuration. Observations
were made in multi-channel 
continuum spectral line mode with two sets of $8 \times 3.125
\MHz$\space channels to minimise chromatic aberration, centred on
intermediate frequencies 1.365\GHz\space and 1.435\GHz. Each
channel was composed of two independent circular polarisations. This
observational arrangement is known as `4IF' mode, in VLA parlance.
These observations were followed in August 1998 by 3 observations of
10hrs each in the A configuration.  These observations were also made
in 4IF mode, centred on the same frequencies as before.

\subsection{Initial Data Reduction}

Both the A and B configuration observations were calibrated with the
same primary and secondary calibrators, 1331+305 (3C286) and
1310+323, for amplitude and phase respectively. As the observations
were made in spectral line mode, bandpass calibrators were also
needed, 1310+323 for the A configuration data and 1331+305 for the B
configuration data. The B configuration data was calibrated using the
standard recipe for spectral line data within the NRAO \aips
software. After removing the sidelobes of bright sources outside the
X-ray survey area from the data, the noise level of the B
configuration map was close to the theoretical thermal noise level.

Of the 24~hours on-source during the A configuration observations,
10~hours were lost, or unusable, mainly due to thunderstorm related
problems which lead to power outages at some of the antennae,
correlator problems and high interference. The remaining 14~hours of
good data were calibrated and mapped in the standard fashion with
\aipsc. However heavy cleaning and careful positioning of \clean boxes 
were not sufficient to remove fully the sidelobes of bright
sources and thus reach the thermal noise limit. It was therefore
necessary to self-calibrate the A configuration data. The B
configuration data did not require self calibration.

\subsection{Self-calibration of the data}
\label{section:problem}

The brightest source near the field is an extended source on the
western edge of the primary beam. This source is known as
J133328.7+375553 in the FIRST radio survey \citep{first} and has a
flux density of $92\mJy$ at $1.4\GHz$.  Hereafter it is referred to 
as source 1, in
Table \ref{tab:20cm}. It has a linear size of $\sim40''$ which is
just beyond the limit of extended structure detectable by the VLA in A
configuration. It is the brightest source near the field by almost an
order of magnitude and its residual sidelobes, after cleaning, were
the major noise contributors to maps of the area within the primary
beam.  The next 3 brightest sources within the primary beam could also
not be deconvolved properly, even after many \clean iterations.
Ripples typical of phase calibration errors were visible in the maps
made with the A configuration data.

Traditionally, self-calibration has been restricted to fields containing one
or more bright sources near the centre. Obtaining the theoretical dynamic 
range 
of wide field data like ours has often been done by imaging the interfering 
sources to create a model which can then be subtracted from the UV 
data set, known as uvsubing. This method was attempted very carefully channel 
by channel, but the residual sidelobes across the field remained. The reason 
why this method fails is not entirely clear, but it is likely to be 
linked to the angular 
size of source 1. It was easily bright enough for the extended emission 
to be detected, but the minimum antenna spacings of the VLA A configuration 
meant that it was difficult to obtain a good enough model to uvsub 
especially with its location so close to the edge of the primary beam 
where beam effects change rapidly. 
Recently self-calibration has been applied successfully to 
fields containing only relatively faint sources in the centre, such 
as are common in deep survey work \citep[\eg][and references
therein]{ivison02,smail02}. 
Our field has a handful of sources of flux density 
$\geq 10 \mJy$ within, or close to the edge of, its primary beam and
so is quite satisfactory for self calibration, so here we include all the 
sources in our field, including source 1.

To obtain an input model on which to self-calibrate the data, we first
made $4 \times 4096$ maps with $0.22''$ pixels to cover the full X-ray
field.  We carried out $10^{5}$ \clean iterations and also
included separate \clean boxes around 3 bright sources lying not
far beyond the edge of the primary beam (see Fig.~\ref{fig:20plan}).
All of the \clean components from the above mapping were used as
a model for self calibration. \calib was run twice, iteratively,
in the self-calibration procedure, solving for phase weighted by
amplitude each time. After each run of \calib a new map was made to 
obtain a new input model. For the first run the solution interval was
60mins and for the second run the interval was 20mins, comparable with
the time between observations of the phase calibrator. A \uv range of
$20-500\kl$ was used to emphasise the emission from Fourier components
on scales of $0.4-10''$, where the model best represents the sky
brightness.  Data outside this range were given a relative weight of
0.1 to reduce the contribution from very small and very large scale
structures, which are mostly spurious.  This procedure led to 6.9 per cent of
the A configuration data being flagged out, but the ripples were
almost completely removed.  Prior to self-calibration, the rms noise of
the image in the best parts of the field was $\sim12\uJy$/beam,
but was considerably higher near to bright sources. After 
self-calibration the rms noise level, for the A configuration data,
was $\sim10\uJy$/beam across almost the whole field, with areas
near to bright sources showing the greatest improvement.

\subsection{The Image Construction}
\label{sec:map}

\bfig
\centerline{\psfig{figure=20pos.ps,height=8cm,angle=270}}
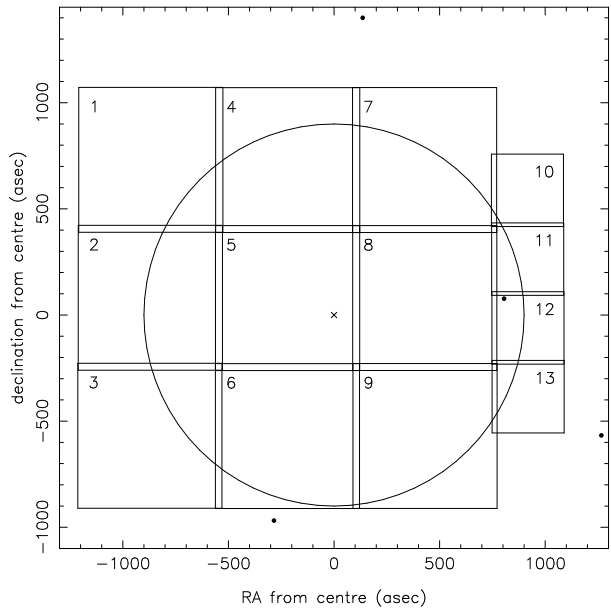
\caption{A figure to show the mapping strategy. Maps 1 to 9 are $2048
\times 2048$ pixels and maps 10 to 13 are $1024 \times 1024$. The
scale is $0.33''$/pixel. The cross in map 5 indicates the centre of
the {\em XMM-Newton} and {\em ROSAT} field of view, and the VLA phase
pointing centre. The small filled circle in map 12 indicates the location 
of the brightest source in the region, known as source 1 in Table A1. The 
locations of three other confusing sources are also indicated by small 
filled circles.}
\label{fig:20plan}
\efig

The A and B configuration data were then combined using the \aips algorithm
\dbconc. The relative weighting of the data sets was determined from
the `sum of the gridding weights' reported by the \aips task
\imagrc, leading to the \dbcon parameter {\sc reweight} = 1, 0.868445.  
A multi-faceted imaging strategy was used so that no part of the survey 
field was too far from the centre of an individual facet image. 
To produce accurate positions, we covered the $30'$ field of view with a
square array of $9\times 2048$ pixel maps, with $0.33''$/pixel.
This pixel scale adequately samples the A configuration restoring beam of 
$1.4''$. This mapping strategy was used to correct for 
the fact that we are approximating the curvature of the celestial 
sphere to a plane. Due to the computing power available it
was possible to deconvolve each field simultaneously in \imagr and use
the {\sc do3d} option (available in more recent versions of \aipsc) 
which reprojects the coordinates of the Fourier transform plane to the 
centre of each facet. The maps were shifted about
$100''$ west to avoid source 1, which had caused problems in the
calibration. The slice of the circular field of view that was missed
was imaged separately with four $1024 \times 1024$ maps in a vertical
stripe (see Fig.~\ref{fig:20plan}).

The final maps were made using natural weighting of the \uv data, to
obtain the most sensitive maps, which gave a restoring beam for the
combined data of $3.35''$. The final set of 13 maps
(Fig.~\ref{fig:20plan}) were constructed with $10^{6}$ \clean
iterations and a {\sc gain} value of 0.1. All fields, except the
flanking field containing source 1 and the flanking field immediately
above it, had pixel noise that was well fitted by a Gaussian
distribution with an average rms of $7.5\uJy$ /beam, prior to
correction of attenuation by the primary beam.  Source detection (see
Section~\ref{sec:sad}) was performed on the array of maps described
above, but for display purposes only, we have produced a $4096 \times
4096$ pixel map, with pixel size $0.33''$, covering the inner $22.8'$
by $22.8'$ of the field (Fig.~\ref{fig:20cmmap}). This map was produced
with a less detailed cleaning strategy and has an rms noise of
$10\uJy$/beam.

\begin{figure*}
{\bf This figure is available in the full version of the paper at www.2.iap.fr/users/seymour/}

\caption{An inverse greyscale image of the inner $22.8'$ by $22.8'$ of
the $1.4\GHz$ map. This map has a pixel size of $0.33''$, is 4096
pixels square and was made with an effective resolution of $3.35''$, 
for display purposes
only. Note that this image has not been corrected for primary beam
attenuation.}  
\label{fig:20cmmap}
\end{figure*}

\section{The Complete Source Catalogue}

\subsection{Source Extraction}
\label{sec:sad}
The \aips algorithm \sad (Search And Destroy) was used to detect
small, and unresolved sources in the maps by fitting two-dimensional
elliptical Gaussians. This task is poor at determining the flux density 
of extended sources as it tries to split them up into small Gaussians.
The flux densities of extended sources were therefore determined by hand using 
the \aips algorithm {\sc tvstat} which finds the total flux density in an
irregular area defined by the user. However there were only 7 sources,
with peak flux density above our detection threshold, where
\sad failed due to extended morphology. \sad was initially instructed
to search for sources above a three sigma peak flux density (ie
$22.5\uJy$), below the final robust detection limit of $30\uJy$ ($4\sigma$). 
This was done to ensure \sad extracted all sources close to the $4\sigma$ 
detection limit where \sad was found to be not so efficient. 
The search was performed on maps prior to correction of
the primary beam attenuation. Therefore, after correction, the $3\sigma$ 
detection limit would be $22.5\uJy$ at the phase centre, but $45\uJy$
at the edge of the X-ray field, i.e.\ $15'$ from the centre.  Each source 
above $4\sigma$ was checked visually to see if it was spurious 
(\eg\space part of a residual side-lobe) and 12 were discarded for 
this reason. Additionally all the
residual maps produced by \sad were inspected for possible real sources with
a peak flux density $>4 \sigma$ which had been missed by
\sadc. This was done by producing contoured maps of the field with the lowest 
contour at $4\sigma$. This led to only 3 more sources being included, all in 
the $4-5 \sigma$ range, all in areas of above average noise. Although slightly 
subjective this method did not change the final catalogue significantly.
Finally all sources with peak flux density $<4\sigma$ were removed from 
the catalogue.

\subsubsection{The $6''$ Resolution Map}

In order to look for low surface brightness objects in the field a map
was constructed in exactly the same fashion as before except that it
was tapered to have a resolution of $\sim6''$, thereby increasing the
sensitivity to low surface brightness sources. The tapered image had
an rms of $12.5\uJy$ and a natural restoring beam of $6.07''$. The
sources were extracted with \sadc, verified in the same manner as before 
and the list correlated with the
original catalogue to look for new sources. Four new, slightly extended
sources with a peak flux density above $4\sigma$ in the tapered map (ie
$>50\uJy$) were discovered and added to the final catalogue which is
listed in table~\ref{tab:20cm}.

\subsection{Flux Density Correction Factors}
\label{sec:bias}
Here we briefly discuss the main factors which distort the true flux
density of sources. For more details see Richards (2000).

\subsubsection{\sad Bias}
\label{sec:sadbias}

There is a known bias in the \sad elliptical fitting algorithm
\citep{ellipgaus} which at low signal/noise overestimates a source's
angular size due to fluctuations in the noise at the edge of a
source. As the total flux density is calculated from the total angular size
and the peak flux density, which is not affected by this bias, the total flux
density is overestimated. This effect was investigated by making 3 separate
simulations where, in each simulation, 64 fake sources of a given flux
density were inserted into each of the nine residual images left after the
running of \sad (\ie the original maps with all the sources found by
\sad removed). These fake sources were inserted with the task \immod
and then extracted by \sadc, having been visually inspected to check
that no sources were coincident with any artifacts in the residual
maps caused by the initial execution of \sadc. This procedure was
repeated for fake sources with peak flux density between $140\uJy$ and
$21\uJy$, at intervals of $7\uJy$. It was then possible to plot the
average overestimate of the total flux density of a source as a function 
of peak flux density.

Figure~\ref{fig:gain} shows how the fractional overestimate increases
with decreasing flux density. An empirical polynomial curve was fitted to the
relationship and was then used to correct the total flux density of every
source. From the simulations it was also possible to calculate the
fraction of sources that are undetected by \sad as a function of peak
flux density (see Figure~\ref{fig:frac}). This relationship was also 
fitted with an empirical polynomial curve and is later used for calculating
the completeness of the differential source counts
(Section~\ref{sec:logn}).

\par
\bfig
\centerline{\psfig{figure=peakvgain.ps,width=3.3in,angle=270}}
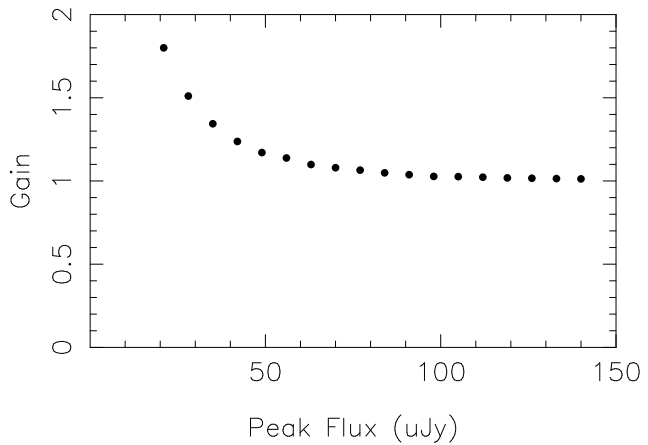
\caption{A plot of the average measured fractional increase in total flux 
density for model point sources inserted into the residual data and then
detected by \sadc, as a function of the input peak flux density.}
\label{fig:gain}
\efig
\par
\par
\bfig
\centerline{\psfig{figure=peakvfrac.ps,width=3.3in,angle=270}}
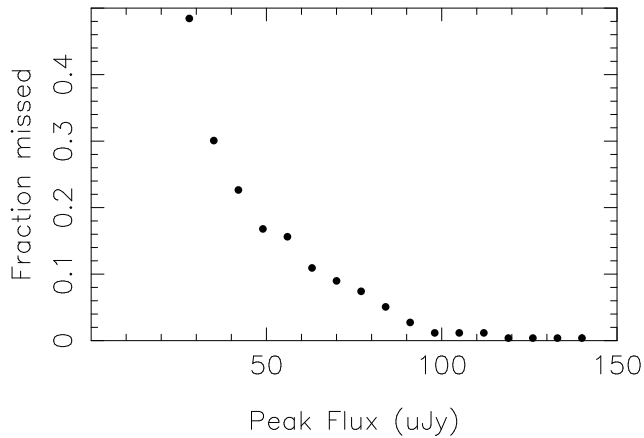
\caption{A plot of the average fraction of sources inserted into the 
residual data which are not detected by \sad shown as a function of
the input peak flux density.}
\label{fig:frac}
\efig
\par

\subsubsection{Instrumental Corrections}
\label{sec:instbias}
There are four instrumental corrections which need to be considered; 
Primary Beam Attenuation, Chromatic Aberration (Bandwidth Smearing), Time Delay
Smearing and 3D Smearing. These corrections are discussed below.

\begin{enumerate}

\item{Primary Beam Attenuation

The ultimate limiting factor to the field of view is the beam 
pattern of the individual antennae. For the VLA at $1.4\GHz$ the 
full width at half power of the beam pattern is $\sim30'$.
The empirically measured primary beam attenuation as a function of 
distance from the pointing centre, as given in the explain file 
for the \aips task \pbcorc, was used to correct the measured 
peak and integrated flux densities of the sources, although the signal to 
noise given in Table \ref{tab:20cm} is the empirically measured 
one which is not affected by the primary beam attenuation.} 

\item{Chromatic Aberration

Chromatic aberration, also known as bandwidth smearing, leads to
sources far from the phase centre being smeared in a radial direction,
with the smearing being greater for larger observing bandwidths. The
peak flux density is less than would be measured with an infinitesimally
narrow observing band, but the total flux density is conserved. Although the
current `4IF' observing set up was selected to minimise the effects of
chromatic aberration, whilst not sacrificing sensitivity, it is still a 
small consideration for sources
far from the field centre.  \cite{hdf} has determined empirically
a relationship, for point sources, between the observed peak flux density 
and the true peak flux density (= the integrated flux density, for point 
sources) for the same `4IF' observing set up that we used. The relationship is

\begin{equation}
  \frac{S_{\rm peak \, flux}}{S_{\rm integrated \, flux}}
=\left\{1+\left(\frac{r}{k}\right)^2\right\}^{-0.5} \\
\end{equation}
\vspace*{5mm}

\noindent where r is the distance from the field centre in arcminutes
and k was found to be $16.19'$ from a least squares fit to the
off-axis measurements of the peak and integrated flux density of the
calibrator source 1400+621.  We have used this formula to correct the
peak flux densities of unresolved sources.  This formula was
also used to estimate the fraction of sources missed in each bin of 
the differential source counts due to the decrease in sensitivity 
to peak flux densities away from the centre.}

\item{Time Delay Smearing

During observations the antennae continuously move through the UV
plane, sampling different parts of the Fourier transform of the sky
brightness. Long integrations therefore lead to averaging of the signal
and consequent distortion of the image. To minimise this `time delay
smearing' we used the minimum 3.3~second sampling interval in all
observations. Using the relationship given in Chapter 18 of Synthesis 
Imaging II \citep{synthesisimaging2} 
at the furthest distance from the phase centre we will have a maximum 
reduction in the peak flux densities of $<0.1\%$ which we did not correct for.
}

\item{3D smearing

A source may be smeared during imaging when the sky is approximated to a 
flat plane. Our mapping strategy to address is discussed in 
section~\ref{sec:map}. To test the positional accuracy of each source
the positions of sources measured from maps made with
large fields of view (\eg\space Fig.~\ref{fig:20cmmap}) were compared
with those from small $30''$ 'postage-stamp' maps. It was found that 
the discrepancies in position were all better than $0.3''$ with no systematic 
offsets as a function of distance from either the pointing centre or the 
map centres. 

Additionally, the positions of non-extended sources were also compared with 
those of potential \rband optical counterparts \citep{chandracat} 
within $1''$ of the radio source position. Again no systematic offsets were 
found. Our paper on the radio/optical correlation (in preparation) will 
explain the matching of optical counterparts in more detail.
No corrections for the annual aberration
effect were made which leads to an discrepancy of almost an order of
magnitude less than the error due to 3D smearing. Total positional
errors are discussed more in Section~\ref{sec:other}.}

\end{enumerate}

\subsection{Catalogue construction}
\label{sec:other}

\subsubsection{Double Sources}
In previous radio surveys a variety of different criteria have been
used to determine whether two nearby unresolved sources are actually
components of one double source, \eg\space a flux ratio $\leq2$ and
separation $\leq2\times$ the restoring beam \citep{ciliegi99}.
\cite{prandoni00} compared the nearest neighbour density pair
distribution function of their catalogue against one expected for a 
random distribution of sources in the sky. They then derived the 
angular distance required to discriminate between real and unreal physical 
associations. None of the sources in 
our catalogue could be considered to be components of a double under the
criteria of \cite{ciliegi99}. If this criteria were relaxed to a flux 
ratio $\leq3$ and a separation $\leq3\times$ the restoring beam we 
obtain 11 pairs. From visual inspection only the two brightest pairs 
appear to be real associations. The first is a single slightly extended 
source associated with a spiral galaxy which \sad has mistakenly 
resolved into two components. The second appears to be two lobes of 
the same source with an optical counterpart between them which we use 
to give the position of the radio source. It is possible that there may be 
further unrecognized doubles in the catalogue but they are likely to be few 
and will not affect the statistics of the source counts.

\subsubsection{Positional Errors}
The systematic error in the position of each sources was taken to be
$0.1''$, which is a conservative estimate for the VLA under normal
conditions and given good phase calibration, such as obtained for these 
observations (VLA Observational Status Summary, \citealp{obssum}). The random
positional error for an unresolved, or only slightly resolved, source is
defined as the point-spread function (\ie\space the FWHM of the
restoring beam, $3.35''$) divided by twice the signal to
noise ratio \citep[see][for a thorough discussion]{ellipgaus}. With a
$4\sigma$ source limit of $30\uJy$ the maximum statistical error is
$0.42''$ giving a maximum total error of $0.43''$.  
We do not give positional errors for individual sources in our
final catalogue (Table 1) but merely note here that the error varies between
$0.42''$ for the faintest sources, down to $0.1''$ for the brightest 
non-extended sources. For the six extended sources mentioned
in section~\ref{sec:sad} the positions of the optical counterparts are
given which are accurate to $0.3''$.

\subsubsection{The Catalogue}
A sample of the catalogue of sources is presented in Appendix A in 
Table~\ref{tab:20cm}. The full Table containing 449 sources above a $4\sigma$ 
peak flux density limit of $30\uJy$/beam appears in the electronic form of
this paper.

\section{The Source Counts}
\label{sec:logn}

Our radio survey was designed primarily to investigate the
relationship between the faint radio and faint X-ray source
populations.  However with an rms noise level of $7.5\uJy$/beam it
reaches approximately the same depth as the deepest radio surveys yet
made at $1.4\GHz$, \eg ~$7.5\uJy$/beam rms noise for the HDF, 
\citep{hdf}; $4.8$ and $9.2\uJy$/beam rms noise for the Lockman Hole East 
and the ELAIS N2 surveys, \citep{ivison02}; $<10\mu$Jy for the Phoenix 
Deep Survey, \citealp{hopkins03}. Our survey is useful in its own right 
for studying the radio source counts at faint flux densities, where
other surveys have shown an upturn, generally attributed to an
additional contribution from starburst galaxies.  In this section we
derive the source counts from our present survey, compare them with
those from other deep surveys and model the counts.

\subsection{Completeness}

In order to minimise the corrections for incompleteness, which become
large at the lowest flux densities, we restrict our sample to sources which
have a peak flux density $>5 \sigma$, \ie $37.5\uJy$ (before correcting 
for the primary beam attenuation).

Although we use a $>5 \sigma$ peak detection threshold so that the
number of false detections will be negligible, our source counts are
based on total flux density. It is therefore possible that some very extended
sources of low surface brightness may be missed. We, as \cite{hdf}, have
searched for such sources using tapered maps, but have found very
few. As the number of additional sources found when going from a
$3.3''$ to $6''$ beam was very small, the number to be found
when searching for even more extended sources is likely to also be very
small, and so we do not consider them further. 

In order to derive the source count density, we simply added up the
number of sources in various total flux density bins of equal width in log 
space: $\Delta S = 0.18 \log(\mJy)$. The exception was the highest flux 
density bin where the upper limit was increased from $682\uJy$ to $1365\uJy$ 
to include all the remaining objects bar the 12 brightest where 
the flux density range greater than $1365\mJy$ is not well sampled. 
The centre 
of each bin was calculated using equation 19 of \cite{wind84} which is a 
function of the normalised differential source count slope (as directly 
measured from our data). Initially each bin centre was simply the mean 
of the flux density of the sources in each bin, but an iterative process 
was used to find the true bin centre by fitting the source count 
slope to our data after deriving new bin 
centres from this equation. After 3 iterations the slope stabilised at 
$0.14\pm0.05$, consistent with that found for the HDF over this region, 
$0.12\pm0.13$ \citep{hdf} (although with a higher normalisation as discussed 
in section~\ref{sec:com} and ~\ref{sec:var}). The number of 
sources in each bin was corrected to take into 
account the number of sources missed by the source extraction
algorithm, \sadc, as explained in Section~\ref{sec:sadbias}. 
The effective area, relative to the full $30'$ X-ray
survey area, was corrected to take into account the attenuation caused
by the primary beam and by chromatic aberration, which increases
radially away from the phase centre. The total correction for these
effects for each bin is shown in Table~\ref{tab:counts} in which we 
also present our final source counts. 

\begin{table}
\caption{The $1.4\GHz$ Source Counts. The first column shows the bin 
range, the second one the bin centre (as explained in the text), the third 
column the number of sources found in that range, the fourth column shows the
multiplicative correction for the biases described in the text and the
final column shows the  source counts per steradian, normalised
to the Euclidean value, for each
bin with Poisson errors derived from the number of sources in each
bin.}
\centering
\begin{minipage}{130mm}
  \label{tab:counts}
  \begin{tabular}{@{}crccc@{}}
    Bin             & $<S_{1.4}>$ & $N_{s}$ & Correction & $S^{2.5}\frac{dN}{dS}$ \\
        ($\uJy$)        & $(\uJy)$    &               &            & $(\Jy^{1.5}{\rm sr^{-1}})$ \\
&&&& \\
  37.50 -  56.76&  46.75&  94& 2.67& $3.26\pm 0.34$\\
  56.76 -  85.91&  70.76& 126& 1.23& $3.74\pm 0.33$\\
  85.91 - 130.03& 107.10&  92& 1.00& $4.15\pm 0.43$\\
 130.03 - 196.80& 162.11&  48& 1.00& $4.02\pm 0.58$\\
 196.80 - 297.87& 245.36&  25& 1.00& $3.90\pm 0.78$\\
 297.87 - 450.85& 371.37&  16& 1.00& $4.65\pm 1.16$\\
 450.85 - 1365.00& 861.40&  13& 1.00& $5.18\pm 1.44$\\
\end{tabular}
\end{minipage}
\end{table}

\nocite{hdf}
\nocite{pdfs}
\nocite{first}

\begin{figure*}
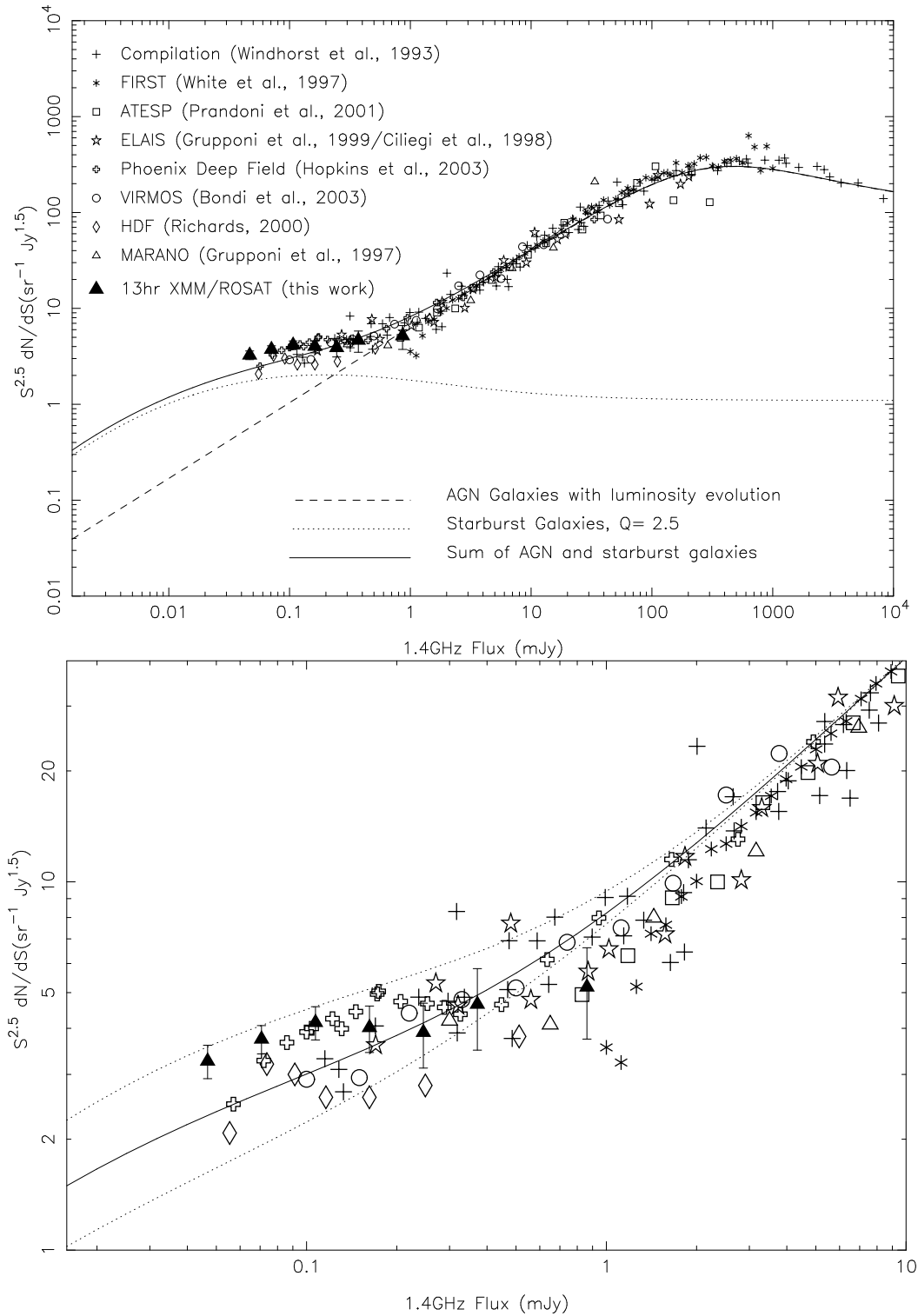

  \begin{minipage}[!]{15cm}
    \centerline{\hbox{
        \psfig{file=final_log.ps,height=10cm,angle=270}
    }}
  \end{minipage}
  \begin{minipage}[!]{15cm}
    \centerline{\hbox{
        \psfig{file=final_log2.ps,height=10cm,angle=270}
    }}
  \end{minipage}

\caption{The $1.4\GHz$ Differential Source Counts of this survey compared 
to other surveys, as indicated on the figure.
{\bf Top Panel~} The solid line represents the combination of the modelled source
counts due to AGN with luminosity evolution (dashed line) and the
counts due to starforming galaxies (dotted line). {\bf Bottom Panel~}
A close up of the sub-$10\mJy$ region with the total model represented by a 
solid line with the evolutionary parameter, Q, for the starforming galaxies 
at 2.5. The dotted lines represent $\pm 0.5$ variations on Q.}
\label{fig:loglog}
\end{figure*}

\subsection{Comparison of our differential source counts with other results}
\label{sec:com}
A summary of recent determinations of the differential $1.4\GHz$
source counts is shown in Fig.~\ref{fig:loglog} along with the results 
presented here. The counts are normalised to 
those counts expected from a non-expanding Euclidean universe 
(\ie $dN/dS \propto S^{-2.5}$). The keys to the various surveys are given
on the figure itself.

The surveys presented in Fig~\ref{fig:loglog} show a consistent upturn above
a Euclidean normalised slope below $1\mJy$. The upturn is interpreted as
the presence of a second source population, in addition to the AGN
population which is responsible for the sources at brighter flux densities.
The second population is usually associated with relatively low radio 
luminosity galaxies at moderate redshifts ($z\leq1$, \eg\space 
\citealp{pdfs} and references therein) and the radio emission is
thought to arise from starbursts \citep[c.f.][]{benn93,hsrr85}.

Apart from the upturn, the most noticeable feature of the sub-mJy
counts is the relatively wide variation between different surveys 
below $1\mJy$ where the upturn becomes noticeable 
(see the close up in the lower panel of
Fig.~\ref{fig:loglog})\footnote{The low values of the FIRST survey below 
$2\mJy$ are due to incompleteness as explained by White et al. (1997)}). 
The variation is greatest around $0.1-0.3\mJy$ where the HDF is significantly 
lower than the other surveys. The different surveys seem to re-converge below
$0.1\mJy$ although one should caution that the various flux density correction
factors are greatest at the faintest flux densities (Table \ref{tab:counts}.
Most of the very deep surveys which sample the sub-\mJy\space flux density 
range have similar, relatively small survey areas ($30-40'$ diameter), where 
they are at their most sensitive, although the VIRMOS and full PDF surveys 
cover 1 and 4.5 square degrees respectively.

The difference between the Phoenix and HDF surveys has already been
discussed by \cite{hdf} and \cite{hopkins03}. \cite{georgakakis00}
note that the difference is larger than would be expected 
from the fluctuations associated with galaxy clustering. The
differences therefore probably arise from differing selection
criteria.  As is clear from the presence of the bright radio source on
the western edge of our field, our own survey area had no radio
selection criteria! The only selection criterion was that galactic
obscuration should be as low as we could find in the sky, $N_{H} =
6.5\times10^{19}\cm^{-2}$, confirmed by measurements of $100\um$
cirrus \citep{imh98}. By contrast, the HDF survey was chosen to
contain no radio sources $>1\mJy$ at $8.3\GHz$ \citep{williams}. 
Minimising the number of bright sources in a field may also 
decreases the number of faint sources in the field as it is likely 
that bright and faint sources are similarly clustered to a reasonable 
extent\footnote{\cite{overzier03} found that fainter sources were more weakly 
clustered than brighter sources in the NVSS and FIRST surveys (which are 
dominated by AGN) but below $5\mJy$ it was found that the clustering of 
AGN remains consistent \citep{wilman03}. The clustering of the radio faint 
starforming population remains unexplored, but may be more weakly clustered 
than AGNs below $\sim0.2\mJy$ \citep{wilman03}.}. As the HDF was also chosen 
to be free of bright sources at other wavelengths, particularly stars and
bright nearby clusters of galaxies, and as sources which are bright at
one wavelength are more likely to be bright at another, there are
additional reasons why the HDF should be deficient in radio sources.
The Phoenix Survey was also chosen to be free of bright radio sources, 
but with a much less stringent criteria than the HDF 
($>100\mJy$ at $4.85\GHz$). Our source counts (not selected to
avoid bright sources) lie above the HDF counts and are consistent with 
the VIRMOS and PDF data from $0.2-1.0\mJy$.

If the apparent differences between the HDF and other surveys at 
$\sim200\mu$Jy are confirmed by deeper surveys
still, then the two-component explanation of the source counts might
account for the differences. For example, if the bump in the source counts is
associated with emission from intermediate redshift ($\sim
0.1-0.5$) starburst galaxies, but distant AGN again dominate at the
very faintest flux densities, then the greater structure in the universe
at intermediate redshifts might account for the variance in the source
counts. The effect of large scale structure on faint source counts at these 
levels warrants further analysis, but detailed analysis is beyond the scope 
of this paper. However the possibility of this effect is discussed briefly 
in the following section on modelling.

\subsection{Modelling the source counts}

Radio source counts have, from the earliest times, been used to
examine the evolution over cosmological time of radio sources
\citep{longair66,rr70}. It has been assumed that the evolution of the 
$1.4\GHz$ luminosity function at a given redshift, $\phi_{z}$, 
can be deconvolved into two components: luminosity and
density evolution, parameterized thus \citep{condon84a,saunders90,rr93,pdfs}:

\begin{equation}
  \phi_{z}(L)=g(z)\phi_{0}\left(\frac{L}{f(z)}\right) \\
\end{equation}

\noindent where $g(z)$ and $f(z)$ represent the density and luminosity
evolution respectively and $\phi_{0}$ is the luminosity function at $z=0$.

The sub-$\mJy$ upturn in the normalised differential source counts is
usually modelled assuming a contribution from starforming galaxies,
for which there is increasing evidence \citep{benn93,gruppioni03},
in addition to that from the AGN \citep{condon84a,rr93,pdfs}
which are associated with the most luminous radio sources
\citep{dunpeak90}. 
Rowan-Robinson et al. refer to the starforming galaxies (SFG) and AGN
as spirals and ellipticals respectively, as the radio AGN are
typically associated with early type galaxies and radio emission from
spirals is normally associated with starformation.
Here we fit the two-component model to our source counts. 

To derive the flux density distribution of sources of any type, AGN or
starburst, one can integrate the local luminosity function using a
parameterisation of the evolution of sources and the chosen cosmology
(throughout we use the cosmological parameters resulting from the WMAP
observations
\citep{wmap} i.e : $\Omega_{M}=0.27$, $\Omega_{\Lambda}=0.73$ and 
$H_0=70\kmpspMpc$).

\begin{equation}
  dN(S) = \int \phi_{z}(L)\thinspace d(\log_{10}L)\thinspace dV \\
\end{equation}

\noindent Density evolution is normally expressed as a
function of redshift, $g(z)\propto(1+z)^P$. Therefore the number of 
objects within a certain flux density range can be
expressed as a function of the flux density, $S$, and the redshift, $z$:

\begin{equation}
  dN(S) = \int (1+z)^{P} \phi_{0}\left(\frac{L(S,z)}{f(z)}\right)  \frac{dS}{S} dV_{c}(z) \\
\end{equation}

\noindent where 

\begin{equation}
  L(S,z)=4\pi\thinspace S D_{L}^{2}(1+z)^{-\alpha-1}.
\end{equation}

\noindent
$D_L$ is the luminosity distance, $\alpha$ is the spectral index 
of the radio source ($S\propto\nu^{\alpha}$). $dV_{c}(z)$ is the comoving 
volume element of solid angle $d\Omega$ and redshift interval $dz$ 
defined by \cite{hogg99} thus:

\begin{equation}
  dV_{c}(z)=D_{H}\frac{(1+z)^{-2} D_{L}^{2}}{(\Omega_{M}(1+z)^{3}+\Omega_{k}(1+z)^{2}+\Omega_{\Lambda})^\frac{1}{2}}d\Omega dz
\end{equation}

\begin{equation}
  \ie dV_{c}(z) = h(z) d\Omega dz
\end{equation}

\noindent where $D_{H}$ is the Hubble distance ($c/H_{0}$), $\Omega_{M}$ 
matter density parameter, 
$\Omega_{\Lambda}$ is the $\Lambda$ density parameter and $\Omega_{k}$ is the 
curvature term. Normalising to a non-expanding, Euclidean universe (\ie 
$dN/dS \propto S^{-2.5}$):

\begin{equation}
  S^{2.5}\frac{dN}{dS}\propto S^{1.5}\int_0^{\infty}\phi_{0}(S,z) h(z) (1+z)^{P} dz \\
\end{equation}

\subsubsection{AGN}
\label{sec:AGN}

\noindent The AGN luminosity function is taken from \cite{dunpeak90},
and shifted from $2.7\GHz$ to $1.4\GHz$ by \cite{rr93}.
Following \cite{rr93} we assume pure luminosity evolution and
no density evolution.  The sample consists of two components, a steep 
spectrum and a flat spectrum component.

Steep spectrum component:

\begin{equation}
  \phi_{0}\left(\frac{P_{1.4}}{P_c(z)}\right)=10^{-6.91}\left[\left(\frac{P_{1.4}}{P_c(z)}\right)^{0.69}+\left(\frac{P_{1.4}}{P_c(z)}\right)^{2.17}\right]^{-1} \\
\end{equation}

\noindent where the luminosity evolution ($P_c(z)=f(z)$) is described by

\begin{equation}
  \log P_c(z) = 26.22 + 1.26z -0.26z^2 \\
\end{equation}

Flat spectrum component:

\begin{equation}
  \phi_{0}\left(\frac{P_{1.4}}{P_c(z)}\right)=10^{-8.15}\left[\left(\frac{P_{1.4}}{P_c(z)}\right)^{0.83}+\left(\frac{P_{1.4}}{P_c(z)}\right)^{1.96}\right]^{-1} \\
\end{equation}

where

\begin{equation}
  \log P_c(z) = 26.36 + 1.18z -0.28z^2 \\
\end{equation}

\subsubsection{Starburst Galaxies}
\label{sec:SFG}

The local SFG luminosity (with no evolution) is represented by the form proposed by \cite{saunders90}.

\begin{equation}
\phi_{0}(L)=C\left(\frac{L}{L_*}\right)^{1-\alpha_f}exp\left[-\frac{1}{2\sigma^2}\log^{2}_{10}\left(1+\frac{L}{L_*}\right)\right] \\
\end{equation}

\noindent which has the advantage of behaving as a power law for $L\ll
L_*$ and as a Gaussian in $\log L$ for $L\gg L_*$. Rowan-Robinson
\etal use this form of the local luminosity density function derived
from a complete sample of radio sources identified with optical
spectra with a flux density above $0.1\mJy$ \citep{benn93}. \cite{sadler02}
fit the same form to data obtained from the cross correlation of the
$1.4\GHz$ NRAO VLA Sky Survey (NVSS) with the 2dF Galaxy Redshift
Survey (2dFGRS). We used the latter fit as although the NVSS has a
higher sensitivity limit than the survey of \cite{benn93} ($2.3\mJy$ 
instead of $0.1\mJy$), the Benn et al. sample is based on only 87 sources 
(56 identified as SFGs) whereas \cite{sadler02} have a final sample of 757 
radio sources (242 identified as SFGs) and hence better represent the local 
luminosity distribution of starforming galaxies. The values given for this fit 
by Sadler et al. are 
$\log_{10} C = -2.41 ({\rm mag}^{-1} {\rm Mpc}^{-3} h^3)$ ($h=H_0/100$),
 $\log_{10}L_*=19.55$, $\alpha_f=0.840$ and $\sigma=0.940$ where the observed
flux densities have been $k$-corrected assuming a mean spectral index of 
$\alpha=-0.7$.

\subsubsection{Total Source Counts}

The total model source counts are the sum of the contributions from AGNs and
SFGs.  The AGN local luminosity function, as described in
Section~\ref{sec:AGN}, is integrated up to a redshift of 5;
integrating up to higher redshifts does not affect the contribution
of AGNs as they would generally be too faint to be detected in surveys
with limits of $0.01-0.1\mJy$. We do not vary any of the parameters from 
those given in Section~\ref{sec:AGN} and so the AGN contribution to the 
total source counts is fixed at the same level as that described by 
previous researchers.

The global space density of starformation rate of the universe increases 
sharply from the present epoch to $z=2$ \citep{madau96,madau98,lilly96} and 
then is generally considered to be constant up to $z=5$
\citep{steidel99,haarsma00,thompson03}, although there is some disagreement 
as to the exact form of the starformation history of the universe after 
$z\sim1.5$ \citep{lanzetta02}. We, as with \cite{pdfs}, therefore
describe the luminosity evolution of the luminosity function of SFGs
as $(1+z)^{Q}$ from redshift 0 to 2, constant between redshifts
2 and 5. We do not consider starforming galaxies at higher redshifts where 
even the most luminous would be undetected at the detection limit of 
this survey. The only parameter which 
we allow to vary in our fit to the combined source counts is the parameter,
$Q$. As with the AGN, there is no density evolution of the SFGs, ie $P=0$. 
All other SFG parameters are fixed at the values
derived by \cite{sadler02} and described in Section~\ref{sec:SFG}.

Due to the large variance between surveys, we have not performed a
$\chi^{2}$ fit to the source counts as the fit would depend greatly on
which datasets were included. We have simply performed a fit-by-eye to
find a solution which passes approximately through the middle of most of the
data points. The value of $Q$ which we thereby derive is 2.5 and the 
resultant fit is shown in Fig.~\ref{fig:loglog}. In that figure
the SFG contribution is represented by a dotted line, the AGN
contribution as a dashed line and the total source counts as a solid
line. Our source counts are represented by solid triangles and several
other surveys which reach below $1\mJy$ are also shown.
In Fig.~\ref{fig:loglog} we also illustrate the fits obtained with
$Q=2.0$ and $Q=3.0$, which lie towards the outer bounds of the distribution
of source count points and so provide some estimate of the error on $Q$.

We note that our best estimate of $Q$ deviates only very slightly from the
values obtained by \cite{pdfs} of $Q=3.1\pm0.8$, \cite{rr93} of
$Q=3.3\pm1.0$, from the \IRAS observations of \cite{saunders90} 
($Q=3.2\pm0.1$), which are believed to be tracing the same starforming 
galaxies, and the value of \cite{condon02}, $Q=3.0\pm1$.
Any differences probably result from the fact that we used the Sadler et al. 
luminosity function for starforming galaxies, our choice of a different 
cosmology to that of previous authors does significantly change the value 
of $Q$.

Although a two component AGN/SFG model may be over-simplistic as AGN
and starforming activity often occur together in the same galaxy
and are probably not entirely independent,
nonetheless it does provide a reasonable explanation of the radio
source counts at faint flux densities. The model is testable as the redshift
distribution of each population can be predicted and compared with
observation when full optical spectroscopic follow-up has been completed. 
In addition, forthcoming {\it Spitzer} \citep{gallagher03} observations of this
field will distinguish SFGs and allow us to refine the
the radio/IR correlation \citep{carilliyun00,gruppioni03}. 

\subsubsection{Variance Between Surveys}
\label{sec:var}

As we already noted, there is significant variance in the source
density between surveys below 1mJy. The variations are of order
$\sim50\%$ at flux density levels of a few hundred $\mu$Jy. If the major
contributors at that flux density level are starburst galaxies, then we 
expect 1.4GHz radio luminosities of $\sim10^{22}$ to $10^{24}$ W Hz$^{-1}$
\citep{gruppioni03,garrett02}. A galaxy of luminosity 
$\sim10^{23}$ W Hz$^{-1}$, with a flux density of $200\mu$Jy, would lie at a
redshift of $\sim0.4$.  At this redshift, filamentary structure in the
universe is already very well developed and may account for the
variance between surveys.  For example, at a redshift of 0.4, the
angular size of our survey corresponds to $\sim9$Mpc. The largest 
structures known locally (\eg\space the Shapley supercluster) are around 
$\sim30$Mpc \citep{quintana00}. To view this volume at a redshift 
$z\sim0.4$ an area of diameter $\sim2^\circ$ needs to studied. There 
is also evidence of very large structures in the radio of $\sim100$Mpc 
below redshift 
$z\sim0.4$ \citep{brand03}. To allow for structure on this scale would 
require surveys of diameter $\sim5^\circ$ at sensitivities below $0.1\uJy$, 
larger than even the largest survey of this depth, the {\it Spitzer} First 
Look Survey radio catalogue \citep[\eg][]{condon03} which has a diameter of 
$\sim3^\circ$.

Additionally, Bellanger \& de Lapparent\nocite{bellap95} (1995 - see 
also Frith et al 2003\nocite{frith03}) show that, at such a
redshift, and locally, the Universe consists of filamentary structure
(walls) separated by great voids. We may see evidence of one such wall
in our own survey as a number of identifications of the {\em ROSAT}
sources lie at a similar redshift of 0.38.  The walls are of size
$\sim100$Mpc and voids have diameters of a few tens of Mpc. Our
survey area is typical of other deep radio surveys and so, depending
on whether the survey area happens to hit a large number of walls, or
pass mainly through voids, large variations in source numbers may be
expected. The Phoenix Deep Survey also appears to show the same kind of
redshift clustering, suggestive of these features of large
scale structure (Andrew Hopkins, private communication). No doubt 
our understanding of large scale radio structure at $z\sim0.4$ will improve 
with detailed analysis of the current surveys mentioned here and others.

\section{Conclusions}

From a total of $\sim 25$hours of data from the VLA at $1.4\GHz$ in
both A and B configuration, 449 sources are detected down to a peak
flux density $4\sigma$ limit of $30\uJy$ in a circular field of $30'$
diameter. This survey is one of the most sensitive  
made at this frequency. 

The source counts are similar to that of other deep surveys, showing an
upturn of the Euclidean normalised source counts below $1\mJy$ and 
are similar to those of the deepest Phoenix sample \citep{hopkins03},
and both lie above the counts from the HDF. The HDF may be
underdense on account of its selection criteria.

As in other deep surveys we assume that
the upturn of the source counts results from a
starforming population at medium redshifts. We base our modelling of
this population on the local galaxy luminosity function of Sadler et
al. (2002), together with mild luminosity evolution. Although our
source counts are similar to those of \cite{hopkins03}, 
our modelling results in a slightly different value of the
luminosity evolution parameter, $Q$, probably due to our choice of the 
Sadler et al. local luminosity function (the choice of a different cosmology 
to that of previous authors does not strongly affect this value). 
Future guaranteed {\it Spitzer} 
observations of this field will help distinguish the SFG and AGN 
contributions and refine the two component model of the source counts.

\section*{Acknowledgments}

We would like to thank the referee, Andrew Hopkins, for very constructive 
comments particularly with the modelling. Also we thank the staff of the VLA 
for providing the data described here, and Tom Muxlow for help in calibrating
the data. NS acknowledges receipt of a PPARC studentship and 
support by the \emph{Probing\- the Origin\- of the\- Extragalactic\- 
background (POE)\/}, European Network number HPRN-CT-2000-00138. IMcH
acknowledges the support of a PPARC Senior Research Fellowship, and
KFG was supported by PPARC grant PPA/G/S/1999/00102.

\bibliography{paper.bbl}

\appendix

\section{The $1.4\GHz$ source catalogue}

Here we present a sample of the complete $1.4\GHz$ source catalogue. The full 
catalogue appears in the electronic online version of this paper. Sources are
ordered by total flux density and the first column gives a source's position
in the catalogue. Columns 2,3 and 4 are the Right Ascension (hrs min secs)
and columns 5,6 and 7 are the Declination (degrees mins secs). Column
8 is the mean angular size of a source (zero means it is
unresolved). Column 9 is the signal to noise ratio before corrections
are made for the attenuation of the primary beam.  Columns 10, 11 and
12 are the peak flux density, integrated flux density and the error of the 
total flux density all corrected for the primary bream attenuation and 
bandwidth smearing.

\begin{table*}
 \centering
 \begin{minipage}{140mm}
  \caption{Complete $1.4\GHz$ Source Catalogue (source number, RA, dec, mean angular size, SNR, peak flux density, integrated flux density and error).}
  \label{tab:20cm}
  \begin{tabular}{@{}lllrrrrr@{}}
   Source   &Right Ascension&Declination      &$\theta_{dec}$&$\sigma_{snr}$ &$S_{\rm peak}$ &$S_{\rm integrated}$&$\Delta S$ \\
        &hh  mm ss      &$^{\circ}$$'$$''$&$        ('')$&               &$(\uJy)$   &$(\uJy)$        &$(\uJy)$   \\
  1      &13 33 29.040   &37 55 57.90       &35.0          &1117.9        &13985.2  &92681.0  &256.5  \\
  2      &13 33 21.353   &37 54 15.75       &70.0            &53.0          &400.0  &12554.7  &598.5  \\
  3      &13 34  3.011   &37 59 49.18        &1.8          &1160.7        &10539.6  &11624.0   &20.6  \\
  4      &13 34 38.496   &38 06 27.01        &2.1           &999.9        &10980.8  &10939.0   &23.4  \\
  5      &13 35 16.664   &38 00  8.19        &3.5           &515.7         &4946.5   &8017.0   &26.9  \\
  6      &13 33 59.951   &37 49 11.64       &12.5           &236.3         &2227.9   &3918.8  &498.8  \\
  7      &13 34 57.650   &37 50 29.40       &25.5            &79.4          &653.0   &3680.7    &150.2  \\
  8      &13 34 38.015   &37 57 10.33       &13.5           &109.3          &833.1   &2510.5  &178.6  \\
  9      &13 35  6.603   &38 03 48.64        &2.0           &210.7         &2176.6   &2260.0   &22.0  \\
 10      &13 34 49.260   &38 05 52.50       &27.5            &51.7          &555.8   &2146.9  &108.4  \\
 11      &13 35 25.340   &38 05 33.90       &23.0            &42.0          &569.8   &1790.6  &125.5  \\
 12      &13 34 38.106   &37 41 33.46        &2.3           &150.1         &1836.5   &1715.0   &26.1  \\
 13      &13 35 35.475   &37 53 14.30        &2.1           &124.5         &1359.6   &1362.4   &21.8  \\
 14      &13 34 13.571   &37 45 39.30        &4.3            &56.0          &557.8   &1112.0   &33.8  \\
 15      &13 34  5.407   &38 07 37.99        &3.5            &70.9          &955.0   &1103.2   &38.0  \\
 16      &13 33 46.583   &38 00 22.26        &2.4            &90.5          &974.2   &1041.9   &27.1  \\
\end{tabular}
\end{minipage}
\end{table*}  

\bsp

\label{lastpage}

\end{document}